\documentstyle[epsfig]{mn}
\begin{document}

\title[Cygnus X-1]
{
The very flat radio -- millimetre spectrum of Cygnus X-1
}

\author[R. P. Fender et al.]
{R. P. Fender$^1$\thanks{email : rpf@astro.uva.nl},
 G. G. Pooley$^2$,
 P. Durouchoux$^3$,
 R.P.J. Tilanus$^{4,5}$,
 C. Brocksopp$^6$
\\
$^1$ Astronomical Institute `Anton Pannekoek' and Center for High
Energy Astrophysics, University of Amsterdam, Kruislaan 403, \\
1098 SJ Amsterdam, The Netherlands\\
$^2$ Mullard Radio Astronomy Observatory, Cavendish Laboratory,
Madingley Road, Cambridge CB3 0HE\\
$^3$ C.E. Saclay, DSM, DAPNIA, Service d'Astrophysique, 91191
Gif-sur-Yvette Cedex, France\\
$^4$ Joint Astronomy Centre, 660 N. A'ohoku Pl., Hawaii, 96720, USA\\
$^5$ Netherlands Foundation for Research in Astronomy, P.O. Box 2,
7990 AA Dwingeloo, The Netherlands\\
$^6$ Astronomy Centre, University of Sussex, Falmer, Brighton BN1 9QH, UK\\
}

\maketitle

\LARGE
\normalsize

\begin{abstract}

We present almost-simultaneous detections of Cygnus X-1 in the radio and mm
regimes, obtained during the low/hard X-ray state. The source displays
a flat spectrum between 2 and 220 GHz, with a spectral index $|\alpha|
\leq 0.15 (3\sigma)$.  There is no evidence for either a low- or
high-frequency cut-off, but in the mid-infrared ($\sim 30 \mu$m)
thermal emission from the OB-type companion star becomes dominant.
The integrated luminosity of this flat-spectrum emission in quiescence
is $\geq 2 \times 10^{31}$ erg s$^{-1}$ ($2 \times 10^{24}$
W). Assuming the emission originates in a jet for which non-radiative
(e.g. adiabatic expansion) losses dominate, this is a very
conservative lower limit on the power required to maintain the jet. A
comparison with Cyg X-3 and GRS 1915+105, the other X-ray binaries for
which a flat spectrum at shorter than cm wavelengths has been
observed, shows that the jet in Cyg X-1 is significantly less luminous
and less variable, and is probably our best example to date of a
continuous, steady, outflow from an X-ray binary. The emissive
mechanism reponsible for such a flat spectral component remains
uncertain.  Specifically, we note that the radio--mm spectra observed
from these X-ray binaries are much flatter than those of the
`flat-spectrum' AGN, and that existing models of synchrotron emission
from partially self-absorbed radio cores, which predict a
high-frequency cut-off in the mm regime, are not directly applicable.

\end{abstract}


\section{Introduction}

Cygnus X-1 (V1357\,Cygni, HDE\,226868) is one of the brightest X-ray
binaries and the classical black hole candidate (BHC); the inferred
mass for the compact object is $\sim 7$ M$_{\odot}$ (e.g. Gies \&
Bolton 1986).  The radio emission in the low/hard X-ray state in the
cm-wave band is persistent, and although it varies by a factor of
about 5 the variations are much less spectacular than those of other
well-studied XRBs such as Cygnus\,X-3 and GRS\,1915+105.  The radio
flux is modulated at the 5.6-day orbital period, particularly at
higher frequencies, and has a spectral index ($\alpha = \Delta \log
S_{\nu} / \Delta \log \nu$) close to zero ($|\alpha| < 0.1$ between 2
-- 15 GHz; Pooley, Fender \& Brocksopp 1999). There are also
variations on timescales as short as 1 hour and long-period
fluctuations with an apparent 140-day period.  The radio emission is
also known to change at major state-changes in the X-ray emission
(Hjellming, Gibson \& Owen, 1975), in the sense that it is reduced in
the X-ray high/soft and/or `intermediate' states.  For long periods
both the radio and X-ray emission are relatively stable, with the mean
radio flux density at cm wavelengths in 1996 -- 1998 being about 14
mJy (Pooley et al. 1999). A detailed comparison of short and long-term
behaviour from radio to hard X-rays is presented in Brocksopp et
al. (1999). Stirling, Spencer \& Garrett (1998) and de la Force et
al. (in prep) have resolved the radio emission from Cyg X-1 on
milliarcsecond angular scales, supporting an origin in a spatially
extended outflow or jet. A single previous detection of $10 \pm 3$ mJy
at 250 GHz reported by Altenhoff, Thum \& Wendker (1994), which was
not simultaneous with radio observations, implied the presence of an
approximately flat spectrum from cm through mm wavelengths.

In this paper we describe multiple detections of the source
simultaneously at cm and mm wavelengths.  These observations reveal a
very flat spectrum extending from the radio to the mm regimes.  They
do not reveal any high-frequency cut-off in the spectrum and raise the
inferred quiescent cm--mm luminosity by a factor of about 15.

\section{Observations}

\subsection{Radio}

Cygnus X-1 is monitored approximately daily by the Green Bank
Inteferometer (GBI) at 2.3 \& 8.3 GHz and by the Ryle Telescope (RT)
at 15 GHz. Details of typical GBI observations can be found in
e.g. Waltman et al. (1994). The RT observations of Cyg X-1 are
described in Pooley et al. (1999).

\subsection{IRAM}

The observations were carried out with the IRAM 30m telescope at Pico
Veleta (Spain) on 1997 Aug 4.  We observed Cyg X-1 simultaneously in
the HCN (J=1-0) transition at 88 GHz, with the SIS 3mm-1 receiver, in
CN (N=1-0) transition at 113 GHz, with the SIS 3mm-2 receiver, in CN
(N=2-1) transition at 220 GHz using the R.230G1 (1mm) receiver.  The
half power beam widths (HPBW) are respectively 27", 22", 10.5" at 88
GHz, 113 GHz and 226 GHz.  The observations were taken using a
position switching mode, with a reference position at -1 arc min in RA
compared to the source position.  The mean system temperatures were
185K, 403K and 514K, and the opacities 0.1, 0.28 and 0.51 respectively
for the 3mm-1, 3mm-2 and 230G1 receivers. Due to the factor 3 between
the opacity at 88 Ghz and 113 GHz, we used the 3mm-1 receiver, with a
512 MHz bandwidth, centered on the HCN (J=1-0) line to estimate the
Cyg X-1 continuum emission. The calibration was taken on the source
K3-50A, and we observed Cyg X-1 for a real time of 30 min, using 12
subscans of 20 s (10 s on source and 10 s off source).  The continuum
emission from Cyg X-1 was detected at a level of 15.9 mJy (3.2$\sigma$)
assuming a 6 Jy/K ratio at 88.6 GHz

\begin{table}
\centering
\caption{Log of IRAM  observations of Cyg X-1 in 1997 Aug}
\begin{tabular}{ccc}
\hline
Date & Frequency & Flux density \\
(MJD)& (GHz)     & (mJy) \\
\hline
50644 & 89 & $15.9 \pm 4.9$ \\
\hline
\end{tabular}
\end{table}

\vspace*{1cm}

\subsection{JCMT}

JCMT SCUBA (Holland et al., 1999) 1350 and 2000~$\mu$m photometry
observations of Cyg X-1 were carried out in 1998 May. Rather than
employing the arrays these observations used the longer wavelength
single-pixel bolometers positioned around the arrays. The data
reduction was performed in the standard manner using SURF (Jenness,
1998; Jenness \& Lightfoot, 1998). Photometric calibration was
achieved by skydip analysis and photometry of Uranus and/or the
secondary standard CRL2688. The emphasis of the observations was on
`detection' rather than accurate flux determinations; the 1350~$\mu$m
flux densities have an accuracy of about 20\%, the 2000~$\mu$m flux
densities of 25-30\%.  The radio, IRAM and JCMT data are plotted in
Fig 1. Least-squares fitting of single power laws to the data from the
two epochs results in spectral indices of $0.07 \pm 0.04$ and $-0.06
\pm 0.05$ for 1997 August and 1998 May respectively.

\begin{table}
\centering
\caption{Log of JCMT observations of Cyg X-1 in 1998 May}
\begin{tabular}{ccc}
\hline
Date & Frequency & Flux density \\
(MJD)& (GHz)     & (mJy) \\
\hline
50945.7 & 146 & $9.2 \pm 3.5$ \\
50950.8 & 146 & $5.8 \pm 3.2$ \\
50960.7 & 221 & $11.6 \pm 2.1$ \\
50964.5 & 146 & $13.2 \pm 4.4$ \\
50964.5 & 221 & $15.6 \pm 2.5$ \\
\hline
all data & 146 & $8.9 \pm 2.1$ \\
   --    & 221 & $13.3 \pm 1.6$ \\
\hline
\end{tabular}
\end{table}

\subsection{XTE}

Cyg X-1 is monitored up to several times daily in the 2-12 keV band
by the Rossi X-ray Timing Explorer (RXTE) All-Sky Monitor (ASM).
See e.g. Levine et al. (1996) for more details.  The total flux 
measured by individual scans is plotted in the top panels of
Figs. 2. Cyg X-1 was in the low/hard X-ray state throughout the period
of these observations.

\begin{figure}
\leavevmode\epsfig{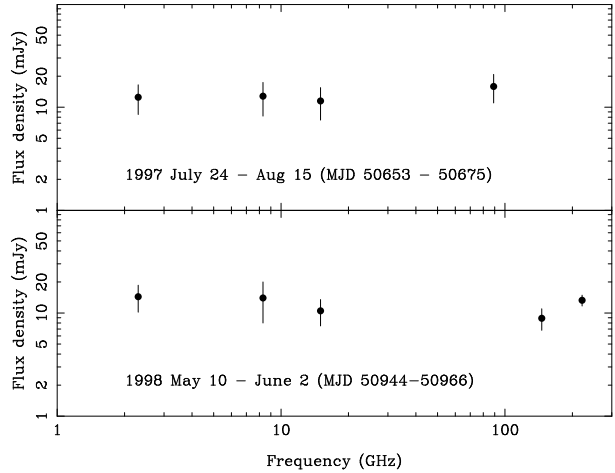}
\caption{Mean 2 -- 220 GHz radio/mm spectra of Cyg X-1 in 1997 August and
1998 May. The spectrum is clearly flat across 2 decades of frequency
at both epochs (best-fit spectral indices of $0.07 \pm 0.04$ and
$-0.06 \pm 0.05$ in 1997 and 1998 respectively).
}
\end{figure}

\section{Discussion}

\subsection{Flux densities at the two epochs}

Table 3 summarises the mean flux densities observed during these two
epochs.  At mm wavelengths (frequencies $\geq 89$ GHz) errors reflect
measurement uncertainties; at cm wavelengths `errors' reflect the
intrinsic source variability, about which little is known at mm
wavelengths. 

\begin{table}
\centering
\caption{Mean 2 -- 220 GHz flux densities of Cyg X-1
in 1997 August and 1998 May.
`Errors' at $\nu \geq 89$ GHz are dominated by measurement uncertainties,
whereas those at 2-15 GHz are dominated by intrinsic source
variability. These data are plotted in Fig 1.
}

\begin{tabular}{lcc}
\hline
Date & Frequency & Flux density \\
(MJD) & (GHz) & (mJy)  \\
\hline
50664 --   & 89    & $15.9 \pm 4.9$ \\
50675    & 15    & $11.5 \pm 4.0$  \\
(1997 July --    & 8.3   & $12.8 \pm 4.6$  \\
August)  & 2.3    & $12.5 \pm 4.0$ \\
\hline
50945 -- & 221 & $13.3 \pm 1.6$  \\
50964    & 146 & $8.9 \pm 2.1$ \\
(1998 May --    & 15  & $10.5 \pm 3.0$ \\
June)     & 8.3 & $14.0 \pm 6.0$ \\
         & 2.3 & $14.4 \pm 4.2$ \\
\hline
\end{tabular}

\end{table}

\subsection{Variability at cm and X-ray wavelengths}

The variations of 15-GHz flux density and X-ray count-rate during the
two observing intervals concerned are shown in Fig 2.  Both intervals
show some enhanced activity in the X-ray band; see discussion in
Brocksopp et al. (1999).  The overall mean amplitude of the 5.6-day
modulation at 15 GHz is 2.0 mJy (zero--peak). Table 3 lists the
r.m.s. variability at 2, 8 \& 15 GHz over a period of 22 days
(i.e. $\sim 4$ orbital periods) centred on each of the two mm
observation periods. In the light of these variations, and the absence
of exactly simultaneous observations, it is possible that the spectrum
in the cm -- mm regime also has some short-term variations, but it is
probable that the mean spectral index is close to zero up to 220\,GHz.

\begin{figure}
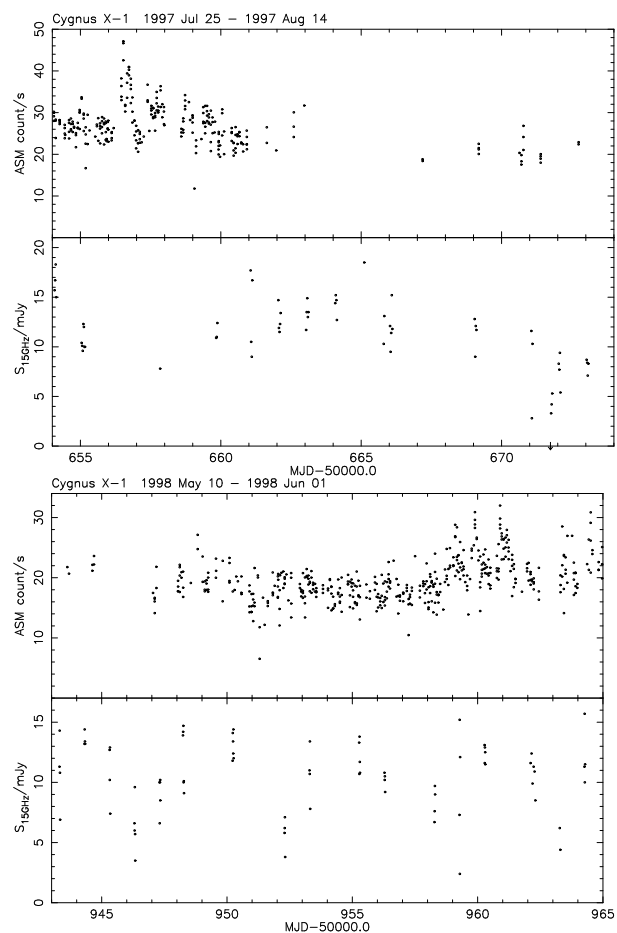

\leavevmode\epsfig{file=97aug-dual.ps, angle=270,width=8cm, clip}
\leavevmode\epsfig{file=98may-dual.ps, angle=270,width=8cm, clip}
\caption{The variations of the 2--12 keV X-ray count rate (upper panels)
and the flux density at 15 GHz (lower panels)
over the two intervals during which mm-wave observation were made.}
\end{figure}

\subsection{The broad band radio -- optical spectrum and a comparison
with other X-ray binaries}

In Fig. 3 we show an extrapolation of the flat radio -- mm
spectrum through the infrared spectral region, combined with published
infrared flux density measurements. It is clear that shortwards of
$\sim 30\mu$m, the observed emission will be dominated by the thermal
component from the OB-type companion star and its wind. Clearly in Cyg
X-1 even if the flat spectrum emission did extend to $\sim 1 \mu$m it
would be extremely hard, probably impossible, to detect, being more
than two orders of magnitude weaker than the thermal emission.

Only two other X-ray binaries have been detected with a flat spectrum
extending from cm to shorter wavelengths.  In GRS 1915+105 the flat
spectrum (inferred synchrotron) oscillations are observed from 13 cm to
$2 \mu$m (e.g. Fender et al. 1997; Fender \& Pooley 1998). In Cyg X-3
the flat spectrum is observed to 0.85 mm (Fender et al. 1995; Ogley et
al. in prep), and maybe also to $2 \mu$m (Fender et al. 1996). Both of
these systems are generally brighter radio sources than Cyg X-1, and
yet also more distant. Table 4 compares the cm-mm luminosities
of the three sources for periods when a flat spectrum is detected
(i.e. Cyg X-1 in the low/hard X-ray state, GRS 1915+105 during periods
of oscillations, Cyg X-3 almost all the time).  It is clear that the
flat-spectrum component in Cyg X-1, as well as being less variable, is
also considerably less luminous than those observed from Cyg X-3 and
GRS 1915+105, by at least an order of magnitude. As noted already in
Fender et al. (1997) the flat-spectrum emission in Cyg X-3 and GRS
1915+105 appears to have approximately the same luminosity. 

GX 339-4 is a persistent black hole candidate X-ray binary with
similar radio properties to Cyg X-1 (Hannikainen et al. 1998; Fender
et al. 1999 and references therein). In particular the source displays
at cm wavelengths a flat spectrum with comparable luminosity to that
of Cyg X-1. We fully expect therefore that sufficiently sensitive
observations should also detect a flat spectrum through the (sub)mm
regime from this source. Additionally, as GX 339-4 is believed to be a
low mass X-ray binary with a less luminous companion star than in the
Cyg X-1 system, we may have more chance of detecting the flat spectrum
at near-infrared wavelengths.

\begin{figure}
\leavevmode\epsfig{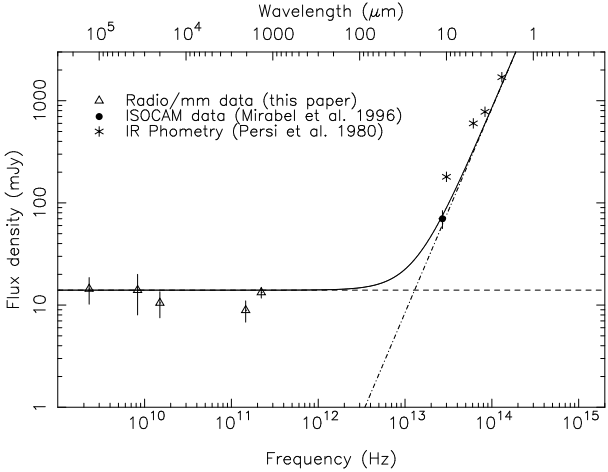}
\caption{The broadband spectrum of Cyg X-1 from radio to optical
wavelengths, combining our radio/mm data with the ISOCAM observations
of Mirabel et al (1996) and ground-based photometry of Persi et
al. (1980). It is clear that even without a high-frequency cut-off
the flat spectral component will be dominated by the thermal
emission from the OB supergiant at wavelengths of 30$\mu$m or shorter.}
\end{figure}

\begin{table}
\centering
\caption{Comparison of cm--mm luminosities of Cyg
X-1, Cyg X-3 and GRS 1915+105}
\begin{tabular}{cccc}
\hline
Source & $S_{\nu}$(flat)  & assumed          & relative \\
       &     (mJy)        & distance (kpc)   & luminosity \\
\hline
Cyg X-1 & 10-15 & 2-3 & 1 \\
Cyg X-3 & 50-100 & 7-10 & 20-250 \\
GRS 1915+105 & 30-60 & 7-12 & 30--215 \\
\hline
\end{tabular}
\end{table}

\subsection{The nature of the flat-spectrum component}

Tables 1--3 and Figs 1 \& 3 summarise the observations of Cyg X-1 for
1997 Aug 4 and 1998 May 11-20. The source is clearly displaying a flat
spectrum through the radio--mm regimes at both epochs. While radio
emission from X-ray binaries is generally assumed to be synchrotron in
origin (see e.g. Hjellming 1988; Hjellming \& Han 1995), in the case
of Cygnus X-1 we do not have direct observational evidence for
this. Even the most rapid variability observed at 15 GHz does not
require a brightness temperature in excess of $10^9$ K, and there is
no direct measurement of linear polarisation. So, while some form of
self-absorbed synchrotron emission remains a possible origin for the
flat spectral component, other emissive mechanisms must also be
considered. 
 
\subsubsection{Energetics}

The observed luminosity of a flat-spectrum source is directly
proportional to the total bandwidth.  In the case of Cyg X-1, the
cm--mm flat spectral component corresponds to a radiative luminosity
of $\geq 2 \times 10^{31}$ erg s$^{-1}$ ($2 \times 10^{24}$ W).
If the emission arises in an
outflow in which non-radiative (e.g. adiabatic expansion) losses
dominate (which seems likely to be the case for relativistic jets from
X-ray transients, see e.g. Hjellming \& Han 1995) then even the
integrated radiative luminosity is only a lower limit on the total
power (i.e. it neglects e.g. electron acceleration and bulk kinetic
energy) required to maintain the jet. Beyond the mm regime, in the
infrared, thermal emission from the companion, stellar wind and
accretion disc begin to dominate the spectrum of the system (Fig. 3)
and it may be very difficult to ever measure any high-frequency limit
to the flat spectral component emission. Note that there is strong
observational evidence that the flat-spectrum oscillations observed
from GRS 1915+105 are dominated by adiabatic expansion losses, based
upon the similarity of the oscillation decay rates at cm and infrared
wavelengths (Fender et al. 1997; Fender \& Pooley 1998).

\subsubsection{Partially self-absorbed synchrotron models developed
for `flat-spectrum' AGN}

It is easy to draw parallels between the flat-spectrum radio--mm
emission from Cyg X-1 (and also Cyg X-3 and GRS 1915+105; see above)
and the `flat-spectrum' extragalactic radio sources. These systems are
generally radio-loud AGN in which the flat-spectrum component
corresponds to the `core' or base of the jet. As pointed out by Cotton
et al. (1980) it would appear to require a `cosmic conspiracy' of
superposition of individual self-absorbed synchrotron components in
order to produce a composite flat spectrum. Marscher \& Gear (1985)
and O'Dell et al. (1988) showed that you can more comfortably
reproduce `flat-spectrum' variability via shocks in conical jets. A
conical jet model for radio emission from X-ray binaries was presented
by Hjellming \& Johnston (1988). Giovanoni \& Kazanas (1990) suggested
that energy transport by relativistic neutrons naturally explained the
combination of electron spectrum, density and magnetic field profiles
required to produce an observed flat synchrotron
spectrum. Alternatively, Wang et al. (1997) have suggested that the
flat-spectrum emission is optically thin from a flattened electron
energy distribution. However, there are problems with the application
of most, possibly all, of these models to the flat
radio--mm(--infrared) spectra observed from Cyg X-1, Cyg X-3 and GRS
1915+105. The simultaneous radio--infrared oscillations observed in
GRS 1915+105 constitute evidence against both shocks which cool via
radiative losses (as the decay rate is the same at 2 cm and 2 $\mu$m)
and an optically thin solution (as the infrared--radio delay, as well
as delays within the radio band, suggest significant optical depth
effects). Furthermore, all the conical jet and related models only
predict a flat spectrum over at most three decades in frequency; the
problem in all cases is the prediction of a high-frequency cut-off
somewhere in the mm band. This is observed in nearly all cases for
`flat-spectrum' AGN, where the mean spectral index in the mm band is
in fact $<\alpha_{\rm mm}> = -0.75 \pm 0.05$ (Bloom et al. 1994). It
is therefore clear that the three X-ray binaries in question have much
flatter (consistent with completely flat) radio--mm(--infrared)
spectra than the `flat-spectrum' AGN, and the applicability of the
self-absorbed synchrotron models to these X-ray binary spectra remains
to be established.

If the emissive mechanism {\em is} synchrotron, then assuming that the
mm emission is not significantly Doppler boosted, we can estimate a
minimum size for the emitting region from the inverse Compton
brightness temperature limit of $10^{12}$K. At 220 GHz, this is only
$10^{10}$ cm, which is relatively close to the compact object and well
within the binary separation ($\sim 10^{12}$ cm) of the system.

\subsubsection{Alternatives to synchrotron emission?}

An obvious candidate for the emissive mechanism of a flat spectral
component is optically thin free-free emission. For optically thin
free-free emission from a thermal plasma, we need to have a
sufficiently large emission measure whilst keeping the spectrum
optically thin to $\nu \leq 2$ GHz. Assuming a fully ionised pure
hydrogen plasma and a Gaunt factor of unity (neither of which
assumptions will affect an order-of-magnitude estimate), and a
distance to the system of 2.5 kpc, we find that we need to satisfy the
following criteria:

\[
r^3 N_e^2 T_e^{-1/2} \geq 4 \times 10^{56}
\]

\noindent
and

\[
r N_e^2 T_e^{-3/2} \leq 2 \times 10^{20}
\]

\noindent
where $r$ is the dimension of the cloud (cm) along the line of sight,
$N_e$ is the electron number density (cm$^{-3}$) and $T_e$ is the
electron temperature (T).  The first criterion is necessary to produce
the observed level of emission, the second to prevent the cloud becoming
optically thick to free-free self-absorption. As a result we can
determine a minimum size of a (spherical) cloud (and corresponding
$N_e$) for different temperatures. For a cloud of $T = 10^4$K, i.e. in
approximate thermal equilibrium with the OB star wind, $r \geq
10^{16}$ cm ($N_e \sim 10^6$ cm$^{-3}$). For a much hotter cloud of
temperature $10^9$K a dimension of $r \geq 10^{14}$ cm ($N_e \sim
10^9$ cm$^{-3}$) is still required. This is very large indeed compared
to the dimensions of the binary orbit, and a significantly larger
emission measure than would be expected for the OB star alone. In this
case the 50\% variability timescale, would be $\geq 1$ hr for a
$10^9$K cloud, and $\geq$ days for $T=10^4$K. The small $(1+4.4 \times
10^{-10}T)$ correction for relativistic free-free emission is
insufficient to significantly alter the result.  Nonthermal optically
thin free-free emission should also produce a flat spectral component,
with (potentially) a greater emissivity than thermal free-free
emission, but precise determination of this (including calculation of
the relevant nonthermal Gaunt factors) is beyond the scope of this
paper. Regardless, as noted above it is difficult to invoke a purely
optically-thin solution as there is evidence for frequency-dependent
delays, indicating a significant optical depth.

Wright \& Barlow (1975) have calculated the spectrum and flux expected
from a spherically symmetric stellar wind as a result of free-free
emission. Combining optically thick and optically thin regimes they
predict a radio--mm spectrum with spectral index $\sim +0.6$. 
The flux density expected from the stellar wind of the OB-type mass
donor in Cyg X-1 (assuming $\dot{M} \sim 2.5 \times 10^{-6}$ and
$v_{\inf} \sim 2000$ km s$^{-1}$) would be around 0.1 mJy at 100
GHz. Therefore we can see that neither the spectrum nor flux density
are compatible with the `standard' spherically symmetric stellar wind
model.

Another possibility is that the radio--mm spectrum is a combination of
some emissive mechanism at radio wavelengths, probably synchrotron,
with a thermal component at (sub-)mm wavelengths. In the case that
this thermal emission arose in an optically thick dust cloud which peaked at a
frequency of $\sim 10^{13}$ Hz ($30 \mu$m), this corresponds to a
temperature of $\sim 150$ K for the dust cloud. At a distance of 2.5
kpc, a spherical cloud of radius $\geq 3 \times 10^{13}$ cm would be
required. This would easily enclose the entire binary system, and
presumably significantly redden the colours of the OB companion star.
In addition, such a large cloud would impose a minimum timescale for
50\% variability of $\geq 10$ min. This cloud size is not unfeasible
for a massive OB-type companion, although in order to be in thermal
equilibrium at $\sim 150$K the dust cloud would need to be much
further from the star (at $10^{13}$ cm from the star the equilibrium
temperature is likely to still be $\geq 1000$K).

\section{Conclusions}

Cyg X-1 was previously known to have a flat radio spectrum from 2 --
15 GHz (Pooley et al. 1999) with a mean flux density of $\sim 14$
mJy. This corresponds to an integrated synchrotron luminosity of $
10^{30}$ erg s$^{-1}$. A single previous observation at 250 GHz had
implied that this flat spectrum extended to mm wavelengths (Altenhoff
et al. 1994). In multiple simultaneous radio and mm observations we
have confirmed the existence of a spectral component extending from cm
through mm wavelengths with a very flat spectrum and no evidence of
either low- or high-frequency cut-offs.  Furthermore, the likelihood
that adiabatic expansion losses dominate in the emitting region shows
that the generation of the outflow may be far more important to the
energetics of accretion in Cyg X-1 than previously
suspected. Presuming the emission to arise in a jet, and comparing
luminosity and variability of this component with that from Cyg X-3
and GRS 1915+105, we infer that the outflow from Cyg X-1 is
considerably steadier and has a significantly lower mass flow
rate. The radio--mm spectra of these X-ray binaries are much flatter
than those of the `flat-spectrum' AGN which generally peak somewhere
in the mm regime and fall off rapidly in the infrared. It is not at
all clear whether models of partially self-absorbed synchrotron
emission from conical jets which have been developed for these AGN can
be extended to apply to the much higher-frequency flat-spectrum
emission we observe from Cyg X-1, Cyg X-3 and GRS
1915+105. 

Detailed spectral measurements in the mm regime, preferably combined
with simultaneous X-ray and radio observations, should help us to
improve the currently inadequate understanding of the emission
mechanisms in this unusual object. Measurement of the shortest
variability timescale (expected to be $\leq 1$ sec for nonthermal
emission with a brightness temperature of $10^{12}$K, $\geq 10$ min
for an optically thick dust cloud at 150 K, or $\geq 1$ hr for
optically thin free-free emission) will be important in understanding
the emissive mechanism. Equally important will be measurement of, or
stringent upper limits to, the level of linear polarisation of the
flat spectral component.  Finally, we predict that a flat spectrum
(sub)mm component will also be detected from the persistent black hole
candidate GX 339-4 in the low/hard state.

\section*{acknowledgements}

We would like to thank the referee for useful suggestions;
additionally RPF would like to thank Kinwah Wu and Robert Voors for
useful conversations.  PD would like to thank Bertrand Lefloch for
assistance with the IRAM observations. IRAM is funded by the Centre
National de la Recherche Scientifique in France, the
Max-Plank-Gesellschaft in Germany and the Instituto Geographico
Nacional in Spain.  The James Clerk Maxwell Telescope is operated by
the Joint Astronomy Centre on behalf of the Particle Physics and
Astronomy Research Council of the United Kingdom, the Netherlands
Organisation for Scientific Research and the National Research Council
of Canada.  We acknowledge with thanks the use of the quick-look X-ray
data provided by the ASM/{\it RXTE} team, and the Green Bank
Interferometer data.  We thank the staff at MRAO for maintenance and
operation of the Ryle Telescope, which is supported by the PPARC.  RPF
is supported by EC Marie Curie Fellowship ERBFMBICT 972436.

\end{document}